\documentclass[footinbib,aps,pra,reprint,superscriptaddress,nopacs]{revtex4-1}

\usepackage{amsmath,amssymb,bbm,bm,graphicx}
\usepackage{csquotes}
\usepackage{amsfonts}
\usepackage{verbatim}
\usepackage{braket}   
\usepackage{mathtools}

\usepackage{amsthm,mathrsfs,amsfonts,epsfig,bm,mathrsfs,enumerate,algorithm,algpseudocode}
\usepackage[shortlabels]{enumitem}

\def\bx{\mathbf{x}}
\def\bz{\mathbf{z}}
\def\bbD{{\bm{\mathcal{D}}}}

\usepackage[absolute]{textpos}
\usepackage{xcolor}

\DeclareMathOperator{\Tr}{Tr}

\begin{document}
\title{A practical and efficient approach for Bayesian quantum state estimation}

\author{Joseph M. Lukens}
\email{lukensjm@ornl.gov}
\affiliation{Quantum Information Science Group, Oak Ridge National Laboratory, Oak Ridge, Tennessee 37831, USA}

\author{Kody J. H. Law}
\email{kodylaw@gmail.com}
\affiliation{School of Mathematics, University of Manchester, Manchester, M13 9PL, UK}

\author{Ajay Jasra}
\affiliation{Computer, Electrical and Mathematical Science and Engineering Division, King Abdullah University of Science and Technology, Thuwal, 23955-6900, KSA}

\author{Pavel Lougovski}
\affiliation{Quantum Information Science Group, Oak Ridge National Laboratory, Oak Ridge, Tennessee 37831, USA}

\date{\today}

\begin{abstract}
Bayesian inference is a powerful paradigm for quantum state tomography, treating uncertainty in meaningful and informative ways. Yet the numerical challenges associated with sampling from complex probability distributions hampers Bayesian tomography in practical settings. In this Article, we introduce an improved, self-contained approach for Bayesian quantum state estimation. Leveraging advances in machine learning and statistics, our formulation relies on highly efficient preconditioned Crank--Nicolson sampling and a  pseudo-likelihood. We theoretically analyze the computational cost, and provide explicit examples of inference for both actual and simulated datasets, illustrating improved performance with respect to existing approaches.
\end{abstract}


\maketitle
\section{Introduction}
Quantum state tomography (QST) is of fundamental importance in quantum information processing, where realization of computational advantages rests critically on the quality of the underlying quantum resources. In general, QST seeks to estimate the density matrix $\rho$ describing a given state, utilizing the results of measurements on repeated state preparations~\cite{Nielsen2000}. As an encapsulation of the quantum state's properties, the density matrix facilitates quantitative predictions of quantum information protocols, clarifies the effects and sources of noise, and provides the foundation for analyzing entire circuits via quantum process tomography~\cite{Chuang1997, Poyatos1997}.

Yet QST is notoriously challenging for all but the smallest quantum systems. The Hilbert space of a collection of qubits grows exponentially with the number of particles, as does the number of independent quantities needed to fully characterize $\rho$. Indeed, such exponential scaling is the source of the unique computational power inherent in quantum information, and accordingly QST cannot be used for characterizing large-scale QIP systems of the future, at least in their entirety. However, there remains demand for efficient and informative QST techniques that make the most of available resources and push limits on system size. In this vein, Bayesian methods offer exciting promise. Built upon Bayes' rule for updating a prior probability distribution according to new information (measurements in the context of quantum tomography), Bayesian QST returns a complete probability distribution on $\rho$, quantifying uncertainty in a natural way, utilizing all available information optimally (in terms of minimizing an operational divergence), and avoiding unjustifiably optimistic estimates of low rank~\cite{Blume2010}. While Bayesian sampling approaches have been applied in several quantum optical experiments~\cite{Williams2017, Lu2019a, Williams2019}, the numerical challenge of drawing from high-dimensional probability distributions impedes widespread use in the physics community.

In this work, we propose, analyze, and demonstrate a full Bayesian tomography method that is straightforward to implement and numerically efficient. Our stand-alone approach leverages recent developments multiple fields, including density matrix parameterization~\cite{Mai2017}, PAC-Bayesian machine learning~\cite{Guedj2019}, and Markov chain Monte Carlo (MCMC) algorithms~\cite{Cotter2013}. After introducing the algorithm in detail, we test it on experimental two-qubit data, obtaining a $\sim$3.5$\times$ speedup in our custom Metropolis--Hastings method over slice sampling. Additionally, with the aid of simulated data of much higher-dimensional two-qudit measurements, we observe a computational scaling advantage utilizing a pseudo-likelihood in favor of a full multinomial likelihood. Overall, our method represents an improvement over previous Bayesian QST approaches and should provide a valuable tool for comprehensive, yet numerically efficient, state estimation.

\section{Background}
In formulating the general problem, consider a system of $n$ qudits---$d$-level quantum information carriers. The Hilbert space dimensionality is then $D=d^n$, and the $D\times D$ density matrix $\rho$ describing a state requires $D^2-1$ real numbers for specification. In order to designate a physically realizable state, $\rho$ must be (i) normalized [$\Tr\rho = 1$], (ii) Hermitian [$\rho^\dagger = \rho$], and (iii) positive semi-definite [$\braket{\psi|\rho|\psi}\geq 0$ for all unit-norm $D$-dimensional states $\ket{\psi}$]. Historically, three major approaches have been adopted to estimate $\rho$ from measurements.

\textit{Linear inversion.---}The first method considered in quantum information processing, linear inversion tomography relies on the fact that measurement outcome probabilities are linear functions of the individual elements comprising $\rho$~\cite{Nielsen2000}. Thus, if a sufficient number of measurements have been performed to access all $D^2-1$ parameters of $\rho$---and the outcome frequencies are equated with these probabilities directly---one can enlist, e.g., least-squares (LS) inversion to obtain an estimate $\rho_{LS}$. While straightforward, LS tends to return nonphysical states: normalization and hermiticity can easily be enforced, but positive semi-definiteness cannot be.

\textit{Maximum likelihood.---}Maximum likelihood estimation (MLE) finds the density matrix which is most likely to have produced the observed data $\bm{\mathcal{D}}$:
\begin{equation}
\label{e1}
\rho_{MLE} = \arg\max_\rho L_\bbD (\rho),
\end{equation}
where $L_\bbD \propto \mathcal{P}(\bbD|\rho)$, the probability of receiving the particular set of outcomes given state $\rho$, as defined by some model~\cite{Hradil1997, James2001}. Through appropriate parameterization of $\rho$, this method guarantees a result satisfying all physicality constraints. This advantage has made MLE the dominant approach to QST in recent years. However, as seen in Eq.~(\ref{e1}), $\rho_{MLE}$ is a point estimate and so does not quantify the level of uncertainty in the result. In practice, error bars have been obtained by modifying the observations according to, e.g., a Poissonian noise model and computing many MLE estimates~\cite{Altepeter2005}, a procedure which amounts to simulating further experiments and averaging the MLE results obtained from these. While likely to give reasonable estimates, this approach is somewhat ad hoc and conceptually undesirable, as it involves feeding in additional data beyond that obtained experimentally.

\textit{Bayesian.---}The third and least explored approach, Bayesian QST~\cite{Blume2010, Granade2016, Williams2017} accounts for experimental uncertainty explicitly through Bayes' theorem. Suppose $\rho(\bx)$ is parameterized by some vector $\bx$, such that any value within $\bx$'s support returns a physical matrix.  Bayes' theorem states that the posterior probability distribution of $\bx$, given results $\bbD$ of some experiment, follows via
\begin{equation}
\label{e2}
\pi(\bx) = \frac{1}{\mathcal{Z}} L_\bbD (\bx) \pi_0(\bx),
\end{equation}
where $L_\bbD (\bx)$ is the likelihood (as in MLE), $\pi_0(\bx)$ is the prior distribution (any beliefs about $\rho$ before the experiment), and $\mathcal{Z}$ is a normalizing constant such that $\int d\bx\,\pi(\bx) = 1$. With access to $\pi(\bx)$, the expectation value of any function $\phi$ of $\rho$ can be obtained
\begin{equation}
\label{e3}
\braket{\phi(\rho)} = \int d\bx\,\pi(\bx)\phi\left(\rho(\bx)\right),
\end{equation}
which can be used to compute, e.g., the mean and standard deviation of any quantify of interest.

Nevertheless, evaluating integrals of the form in Eq.~(\ref{e3}) is numerically challenging due to their generally complicated features and high dimensionality, even for moderate-size systems (e.g., two qubits). Accordingly, MCMC  methods have been invoked in the literature, such as Metropolis--Hastings~\cite{Blume2010, Mai2017}, sequential Monte Carlo (SMC)~\cite{Granade2016} and slice sampling~\cite{Williams2017}. These approaches are designed, in most cases, to obtain $R$ samples $\{\bx^{(1)},\bx^{(2)},...,\bx^{(R)}\}$, so that Eq.~(\ref{e3}) can then be approximated as
\begin{equation}
\label{e4}
\braket{\phi(\rho)} \approx \frac{1}{R} \sum_{r=1}^R \phi\left(\rho(\bx^{(r)})\right).
\end{equation}
Slice sampling in particular is an effective and quite general MCMC method, requiring no proposal distributions and largely insensitive to initial step settings settings~\cite{Neal2003, MacKay2003}. However, the computing time required for convergence can easily make these methods intractable for systems of interest. A major motivation for the current work rests in the realization that a more tailored sampling method---focused on the specific density matrix parameterization and robust to increases in system dimensionality---can attain significant computational speedups.

\begin{figure*}[tb!]
\centering\includegraphics[width=5.5in]{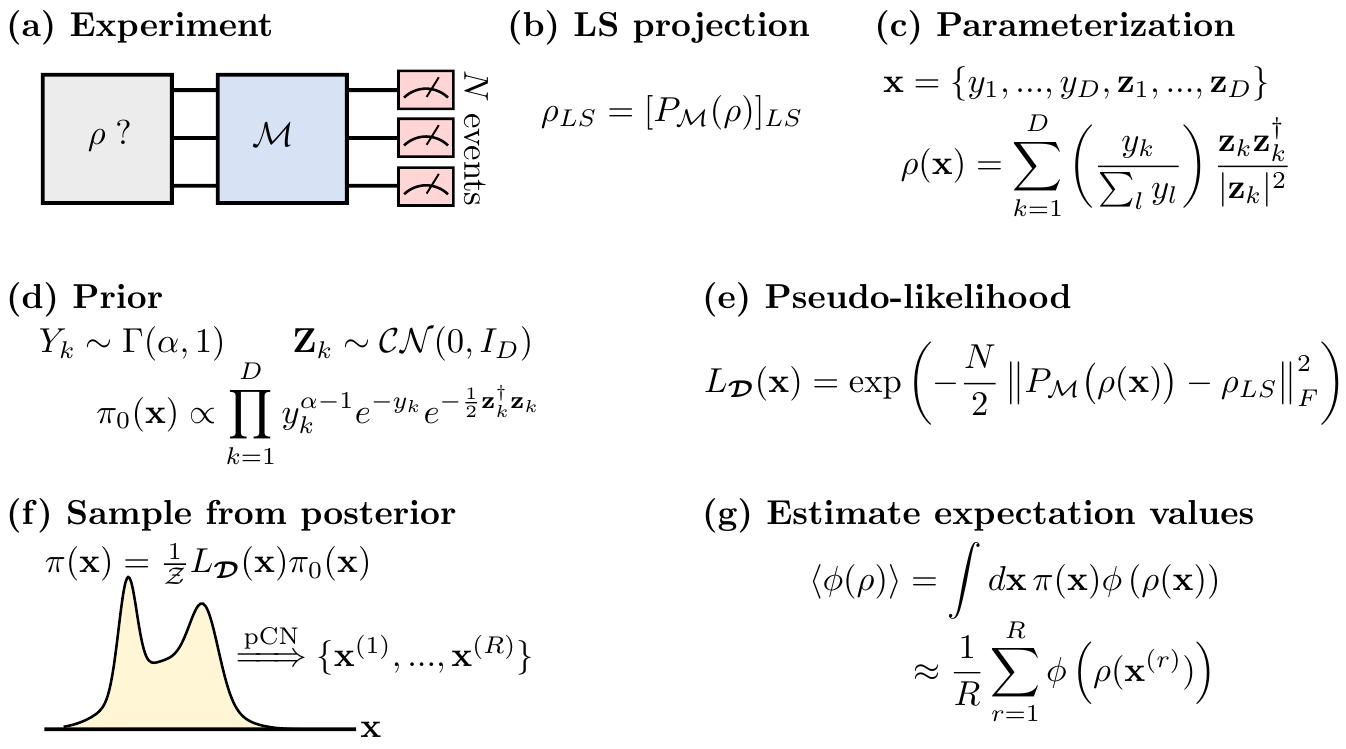}
\caption{Overview of proposed Bayesian QST method.}
\label{fig1}
\end{figure*}

Finally, before describing in detail the procedure introduced here, we note an alternative view of Bayesian tomography associated with adaptive QST. In this application, Bayes' theorem is invoked in real-time, with the results from previous measurements used to hone in subsequent measurement choices and reduce the total number of bases required for reconstruction~\cite{Huszar2012, Kravtsov2013, Struchalin2016, Pogorelov2017, Granade2017}. While beyond the scope of the present work, where we concentrate on Bayesian state reconstruction \emph{post-experiment}, we could certainly envision incorporating aspects of approach into adaptive QST as well.

\section{Proposed Method}
\label{sec:method}
\subsection{Steps}
\label{sec:steps}
We now outline our proposed Bayesian QST workflow, summarized visually in Fig.~\ref{fig1}. (The steps will be explained in detail in Sec.~\ref{sec:details}.)
\begin{enumerate}[(a)]
\item Perform a set of measurements on unknown state $\rho$, amounting to a total of $N$ individual outcomes (over all measurement settings).
\item Compute the least-squares estimate $\rho_{LS}$. If the number of measurements is tomographically incomplete, $\rho_{LS}$ lives in a subspace spanned by only those directions which were observed, which we can express through the function $P_\mathcal{M}(\cdot)$, i.e., $\rho_{LS} = [P_\mathcal{M}(\rho)]_{LS}$.
\item Parameterize the $D\times D$ density matrix  by $D$ nonnegative real numbers, $y_k$, and $D$ complex column vectors of length $D$, $\bz_k$. The density matrix for parameter set $\bx = \{y_1,...,y_D, \bz_1,...,\bz_D\}$ is then
\begin{equation}
\label{e5}
\rho(\bx) = \sum_{k=1}^D \left(\frac{y_k}{\sum_l y_l} \right) \frac{\bz_k \bz_k^\dagger}{\lvert \bz_k \rvert^2}.
\end{equation}
This satisfies all physicality conditions.
\item Take the prior distribution for $\bx$ as
\begin{equation}
\label{e6}
\pi_0(\bx) \propto \displaystyle\prod_{k=1}^D y_k^{\alpha-1} e^{-y_k} e^{-\frac{1}{2} \bz_k^\dagger \bz_k},
\end{equation}
which amounts to treating the weights as Gamma-distributed random variables [$Y_k\stackrel{\textrm{i.i.d.}}{\sim} \Gamma(\alpha,1)$] and the vectors as standard-normal complex Gaussians [$\mathbf{Z}_k\stackrel{\textrm{i.i.d.}}{\sim} \mathcal{CN}(0,I_D)$].
\item Define the pseudo-likelihood as
\begin{equation}
\label{e7}
L_\bbD(\bx) = \exp\left(-\frac{N}{2} \left\lVert P_\mathcal{M}\big(\rho(\bx)\big) - \rho_{LS} \right\rVert_F^2 \right),
\end{equation}
with $\lVert A \rVert_F \equiv \sqrt{{\rm Tr}(A^\dagger A)}$ denoting the Frobenius norm and $P_\mathcal{M}(\cdot)$ the projection introduced in Step (b).
\item Draw $R$ samples according to the preconditioned Crank--Nicolson (pCN) Metropolis--Hastings procedure of invariant distribution $\pi(\bx) \propto L_\bbD(\bx) \pi_0(\bx)$, detailed in Algorithm~\ref{a1}.
\item From these samples, estimate any function of $\rho$ via Eq.~(\ref{e4}).
\end{enumerate}

\begin{figure*}[tb!]
\begin{minipage}{\linewidth}
\begin{algorithm}[H]
\caption{pCN Sampling Procedure}
\label{a1}
\begin{algorithmic}[1]
\State Choose the stepsizes $\beta_y,\beta_z \in(0,1)$. Set $j=0$. and draw $\bx^{(0)}$ from the prior [Eq.~(\ref{e6})]. 
\State For $k\in\{1,...,D\}$, draw $\eta_k\stackrel{\textrm{i.i.d.}}{\sim} \mathcal{N}(0,1)$ and independently $\boldsymbol{\xi}_k \stackrel{\textrm{i.i.d.}}{\sim} \mathcal{CN}(0,I_D)$. Then propose the new point $\bx^{\prime}=\{y_1^\prime,...,y_D^\prime;\bz_1^\prime,...,\bz_D^\prime\}$, according to
\begin{eqnarray*}
y_k^\prime & = & y_k^{(j)} e^{\beta_y\eta_k} \\
\bz_{k}^{\prime} & = & \sqrt{1-\beta_z^2}\bz_{k}^{(j)} + \beta_z\boldsymbol{\xi}_k.
\end{eqnarray*}
\State Set $\bx^{(j+1)}= \bx^\prime$ with probability $A(\bx^\prime,\bx^{(j)})$, where
\begin{equation*}
\log A(\bx^\prime,\bx^{(j)}) = \min\left\{0, \log L_\bbD(\bx^\prime) - \log L_\bbD(\bx^{(j)}) + \sum_{k=1}^D\left[\alpha\log y_k^\prime - y_k^\prime -\alpha\log y_k^{(j)} + y_k^{(j)} \right] \right\}.
\end{equation*}
Otherwise set $\bx^{(j+1)} = \bx^{(j)}$. Increment $j$ by one and return to step 2.
\end{algorithmic}
\end{algorithm}
\end{minipage}
\end{figure*}

\subsection{Further Details on Specific Steps}
\label{sec:details}
\textit{Parameterization.---}We have opted for the parameterization and prior employed by Mai and Alquier~\cite{Mai2017}, which expresses the density matrix as a superposition of normalized (though non-orthogonal) projectors. Incidentally, this represents an over-parameterization, in that it relies on a total of $2D^2+D$ real numbers, rather than the minimum of $D^2-1$ required for a $D\times D$ density matrix. We have found this parameterization significantly more efficient to sample from and evaluate than the Cholesky approach of Refs.~\cite{Seah2015, Williams2017}. For example, computing the determinant in the integration measure of Ref.~\cite{Williams2017}---needed to preserve Haar invariance~\cite{Fyodorov2004}---requires $\mathcal{O}(D^6)$ operations for a given draw. On the other hand, the current over-parameterization utilizes a simple Cartesian differential,
\begin{equation}
\label{e8}
d\bx = \prod_{k=1}^D \left[ dy_k \prod_{l=1}^D d(\mathrm{Re}\,\bz_{k,l}) d(\mathrm{Im}\,\bz_{k,l}) \right],
\end{equation}
where $\bz_{k,m}$ denotes the $m$-th component of the complex vector $\bz_k$. Constructing $\rho$ given $\bx$ requires only $\mathcal{O}(D^3)$ operations [Eq.~(\ref{e5})], offsetting the small overhead incurred from the additional parameters.

The prior distribution [Eq.~(\ref{e6})], also from Ref.~\cite{Mai2017}, is specified by one user-adjustable value, $\alpha$, which can be used to favor low- or high-rank $\rho$, i.e., pure or mixed states, respectively. The collection of $D$ normalized random variables, $Y_k/(\sum_l Y_l)$, with $Y_k\stackrel{\textrm{i.i.d.}}{\sim} \Gamma(\alpha,1)$ follows a Dirichlet distribution $\textrm{Dir}(\alpha)$, which guarantees both normalization and nonnegativity; $\alpha=1$ represents a fully uniform prior, with equal weight given to all physically realizable states, while $\alpha<1$ favors sparse Dirichlet draws~\cite{Telgarsky2013} and hence purer states. (It is important to note that Haar invariance of the prior $\pi_{0}(\bx)$ obtains for any choice of $\alpha$, due to rotational symmetry of the normal distribution.) Finally, the complex Gaussian vectors, when normalized to $\mathbf{Z}_k/|\mathbf{Z}_k|$, correspond to uniform draws from the complex unit hypersphere.

\textit{Pseudo-likelihood.---}The particular $L_\bbD(\bx)$ chosen in Eq.~(\ref{e7}) is ``pseudo'' in that it does not proceed from an explicit experimental model, but rather merely assigns a loss function between a proposed $\rho(\bx)$ and experimental data, in this case the least-squares estimate $\rho_{LS}$. Growing in popularity in the context of ``probably approximately correct'' (PAC) Bayesian machine learning~\cite{Guedj2019}, pseudo-likelihoods are useful when a first-principles model is either unknown or too complex to compute efficiently. The downside is the need to separately specify the weight between evidence and prior, as controlled by the constant appearing in the psuedo-likelihood expression $\exp(\textrm{const} \times \textrm{loss})$. The larger its value, the more sharply peaked around $\rho_{LS}$ the posterior distribution becomes. As one of the strengths of Bayesian QST lies in its quantification of estimator uncertainty, it is essential that this scale factor reflect confidence levels commensurate with the amount of data gathered. For a quadratic loss function as in Fig.~\ref{fig1}(e), one can associate this constant with $1/(2\sigma^2)$, with $\sigma^2$ the variance. If we take $N$ as the total number of events utilized in the LS estimate, it is reasonable to assume $\sigma^2 \propto 1/N$, although the specific proportionality factor is unclear. Reference~\cite{Mai2017} conjectures $\sigma^2 = 2/N$ as optimal, but in the absence of more rigorous motivation, we select $\sigma^2 = 1/N$, which in the examples below leads to uncertainties comparable to that of a full likelihood, albeit slightly larger. 
In general, more thorough methods for selecting the variance represent an important direction for future research.

In Steps (b) and (e), we propose treating cases of incomplete measurements by projecting onto only those elements of $\rho$ which are accessed in the experiment, expressed formally through the function $P_\mathcal{M}(\cdot)$. Consider the decomposition of a $D\times D$ density matrix in terms of $D^2-1$ traceless, Hermitian generators $\lambda_k$ of SU($D$):
\begin{equation}
\label{e8-1}
\rho = \frac{1}{D}I_D + \frac{1}{2} \sum_{k=1}^{D^2-1} c_k \lambda_k.
\end{equation}
In light of orthogonality [$\Tr\lambda_k\lambda_l = 2\delta_{kl}$], we have $c_k = \Tr\rho\lambda_k$. Thus, incomplete measurements reflect that only a subset of the $D^2-1$ observables can be estimated through linear inversion. Suppose that $\mathcal{K}_\mathcal{M}$ denotes this subset of indices; then we define $P_\mathcal{M}(\cdot)$ as
\begin{equation}
\label{e8-2}
P_\mathcal{M}(\rho) = \frac{1}{D}I_D + \frac{1}{2} \sum_{k\in \mathcal{K}_\mathcal{M}} \Tr\{\rho\lambda_k\} \lambda_k.
\end{equation}
For example, in the data utilized in Sec.~\ref{sec:exp}, measurements were sensitive to eight of the fifteen coefficients required to specify a two-qubit state ($|\mathcal{K}_\mathcal{M}|=8$). By contrast, a tomographically complete experiment corresponds to $P_\mathcal{M}(\rho)=\rho$. We emphasize that this particular projector definition is merely a convenient choice, and it differs from the prob-estimator in Ref.~\cite{Mai2017}.

Additionally, even though we have presented a pseudo-likelihood formulation in defining the proposed method, the basic features can be readily applied to a full (model-infused) likelihood as well. Suppose an experiment consists of $Q$ positive-operator valued measures (POVMs) $\Lambda^{(q)}$, each with $S_q$ total outcomes, associated with operators in the set $\Lambda^{(q)} = \{\Lambda^{(q)}_1,...,\Lambda^{(q)}_{S_q}\}$. Then the full likelihood is the multinomial expression
\begin{equation}
\label{e9}
L_\bbD(\bx) = \prod_{q=1}^Q \prod_{s=1}^{S_q} \left[\Tr\rho(\bx)\Lambda^{(q)}_s\right]^{N_s^{(q)}},
\end{equation}
where the outcome associated with $\Lambda_s^{(q)}$ is observed $N_s^{(q)}$ times. Computationally speaking, considering $Q$ measurements, each with $S_q=D$ outcomes (as in standard projective measurement), evaluating this likelihood requires $\mathcal{O}(Q D^3)$ operations. On the other hand, the cost of evaluating the pseudo-likelihood depends on the specifics of the projection operation $P_\mathcal{M}(\cdot)$. For $Q\sim\mathcal{O}(D)$ tomographically complete measurements---such as mutually unbiased bases (MUBs)~\cite{Wootters1989}---$P_\mathcal{M}(\rho)= \rho$ and no projection is necessary, thus leaving a total evaluation cost of $\mathcal{O}(D^2)$ for the pseudo-likelihood, compared to $\mathcal{O}(D^4)$ for the full likelihood. For incomplete measurements, however, $P_\mathcal{M}(\rho)$ must be explicitly computed, thus increasing the pseudo-likelihood's evaluation cost. For example, in the case of $Q\sim\mathcal{O}(D)$ but  not a tomographically complete set, evaluating the pseudo-likelihood can increase to $\mathcal{O}(D^4)$ operations. Thus, the pseudo-likelihood in our formulation is expected to impart a computational speedup for \emph{complete} measurements but not necessarily for the incomplete case, a situation which is consistent with the results of Secs.~\ref{sec:exp} and \ref{sec:sim}.

\textit{Sampling algorithm.---}The most challenging feature of Bayesian methods, sampling the posterior distribution faces slow convergence that becomes arbitrarily slow as dimension increases. In 2013, however, Cotter \emph{et al.}~\cite{Cotter2013} introduced a transformative approach to MCMC sampling which eliminates this ``curse of dimensionality'' for Gaussian priors, under appropriate assumptions on the likelihood. Titled ``preconditioned Crank--Nicolson'' (pCN for short), pCN modifies standard random-walk Metropolis sampling by scaling the previous iteration's position before adding a random shift and generating the proposal $\bx^\prime$. In Algorithm~\ref{a1}, pCN appears specifically in the factor $\sqrt{1-\beta_z^2}$ in Step 2. This small modification simplifies the acceptance probability $A(\bx^\prime,\bx^{(j)})$ significantly with respect to a standard random walk proposal, by removing terms of the form $|\bz^{(j)}|^2 - |\bz^\prime |^2$ from the exponent. The difference in these terms can be large, which necessitates a smaller stepsize to maintain a given acceptance rate. Alternatively, independence sampling from the prior ($\beta_z=1$)  also removes these terms, but the acceptance probability in that case is determined by the ratio of the likelihood at two independent prior samples, which one can expect to be large if the likelihood varies substantially over the support of the prior, i.e., if the posterior differs significantly from the prior. Therefore, unlike both standard random-walk Metropolis and independence sampling, the proposal here preserves random walk behavior \emph{and} provides a simplified acceptance probability.

The specific expression for $A(\bx^\prime,\bx^{(j)})$ follows from the standard form for Metropolis--Hastings~\cite{MacKay2003}. Letting $p(\bx^\prime|\bx^{(j)})$ denote the proposal density, we have
\begin{equation}
A(\bx^\prime,\bx^{(j)}) = \min\left\{1,  \frac{\pi(\bx^\prime)}{\pi(\bx^{(j)})} \frac{p(\bx^{(j)}|\bx^\prime)}{p(\bx^\prime|\bx^{(j)})} \right\}.
\end{equation}
Making use of the densities for the proposal distribution, $Y_k^\prime|\mathbf{X}^{(j)} \sim \textrm{Lognormal}(\log Y_k^{(j)},\beta_y^2)$ and $\mathbf{Z}_k^\prime|\mathbf{X}^{(j)}\sim\mathcal{CN}(\sqrt{1-\beta_z^2}\mathbf{Z}_k^{(j)},\beta_z^2I_D)$, as well as Eqs.~(\ref{e2}), (\ref{e6}), and (\ref{e7}), returns the formula in Step 3. For efficient convergence in the sampling algorithm, we monitor the acceptance rate and increase or decrease the step sizes $\beta_y$ and $\beta_z$ in tandem to maintain an acceptance fraction between 0.1 and 0.3. This range is chosen to enclose 0.234, the optimum acceptance probability, under various assumptions, for random-walk Metropolis--Hastings~\cite{Roberts1997}. Additionally, we note that the adaptation diminishes as the chain evolves, so it does not preclude ergodicity~\cite{Roberts2007}.


\section{Example with Experimental Data}
\label{sec:exp}
To explore the effectiveness of the new method, we perform QST on results from the frequency-bin quantum optics experiment of Ref.~\cite{Lu2018b}, specifically the measurements in Fig.~4 thereof. We selected this experiment for comparison because: (i) its basis set is tomographically incomplete, thus enabling use of the projector formulation in Fig.~\ref{fig1}(b); and (ii) the results were already analyzed with the method of Ref.~\cite{Williams2017}, providing an initial reference point. In the following, we perform numerical benchmarking of method performance for a variety of configurations. All tests were completed in 64-bit \mbox{MATLAB} utilizing a single thread on a 2.5~GHz machine with 128~GB of RAM.

\begin{figure}[tb!]
\centering\includegraphics[width=\columnwidth]{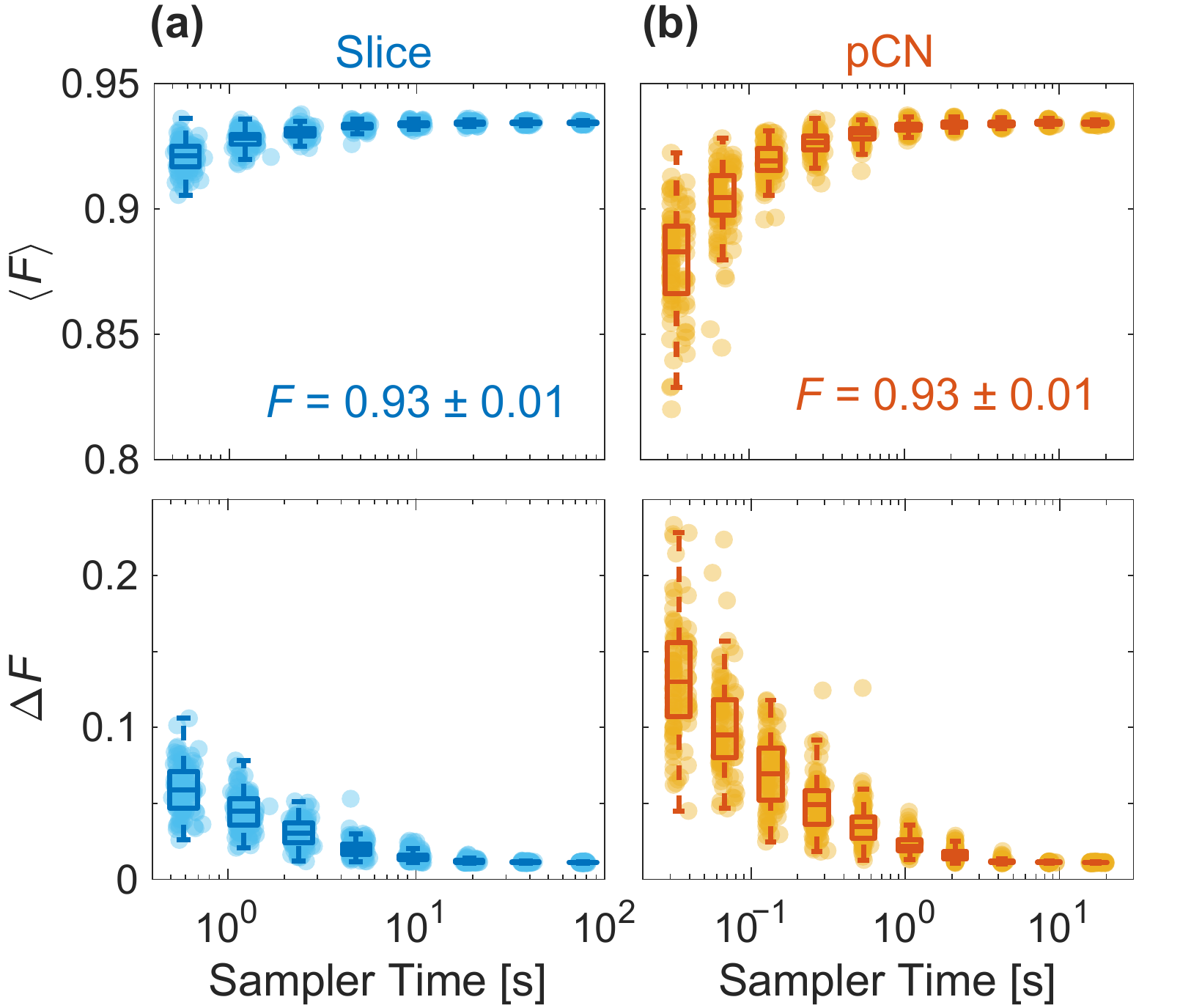}
\caption{State fidelity estimates for the example experimental data for independent samplers as the Markov chain length (and hence total time) increases, for (a) slice sampling and (b) the proposed pCN algorithm.}
\label{fig2}
\end{figure}

For the first test, we focus on the speed of the pCN sampling method, using slice sampling with MATLAB's built-in algorithm for comparison. In this example, we take $\alpha=1$ for a uniform prior and invoke the full likelihood of Eq.~(\ref{e9}). We increase the number of points in the Markov chain and monitor the mean $\braket{F}$ and standard deviation $\Delta F$ of the fidelity $F(\bx) = \braket{\Psi|\rho(\bx)|\Psi}$, where $\ket{\Psi}=\tfrac{1}{\sqrt{2}}(\ket{01}+\ket{10})$ is the ideal entangled state; $R=2^{10}$ samples are kept from a total of $RT$, with $T$ the thinning parameter used to reduce serial correlation in the chain. For each value of $T$, we run 100 independent samplers (i.e., with random initial points), returning  100 separate estimates of $\braket{F}$ and $\Delta F$. The results for slice sampling appear in Fig.~\ref{fig2}(a), plotted against total time logged by each sampler. Thinning increases by factors of two from $2^0$ to $2^7$ on this plot, and we use a box plot format to summarize the statistics at each $T$: the center mark denotes the median, upper and lower lines enclose the the 25th--75th percentiles, and the whiskers extend to the smaller of the farthest point or $1.5\times$ the length of the box. Fidelity converges to $F=0.93\pm0.01$, slightly higher than the mean of 0.92 found in Ref.~\cite{Lu2018b}. This difference is unsurprising, though, since here we do not consider singles counts (i.e., events where only one of the two photons is detected), so our likelihood model differs.

Figure~\ref{fig2}(b) furnishes results for the identical test performed with our custom pCN algorithm, for thinning from $2^0$ to $2^9$. The total times are lower by approximately 18$\times$ for the same value of $T$. Examining the codes in detail indicates this difference is caused by the fact the slice sampler evaluates $\pi(\bx)$ many more times than pCN. Nevertheless, Fig.~\ref{fig2} reveals pCN's need for larger thin values to reach the same level of convergence as slice, so it is not clear a priori what, if any, quantitative advantage is obtained.

\begin{figure}[tb!]
\centering\includegraphics[width=\columnwidth]{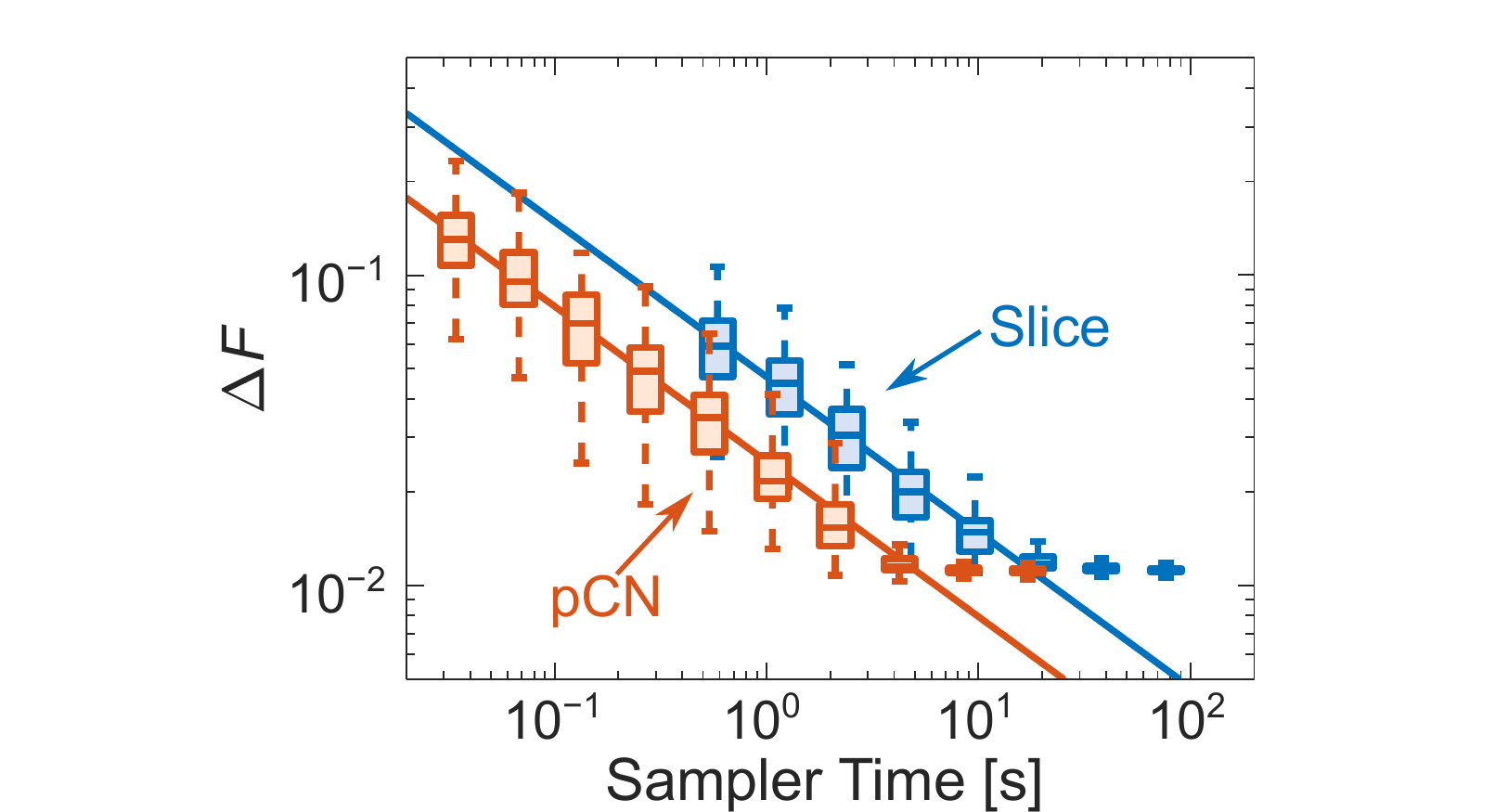}
\caption{Convergence of $\Delta F$ in the results of Fig.~\ref{fig2}. The fits have log-log slopes of $-1/2$, with the pCN curve shifted to the left by a factor of $\sim$3.5$\times$.}
\label{fig3}
\end{figure}

Accordingly, we next plot $\Delta F$ for both slice and pCN on the same logarithmic scale in Fig.~\ref{fig3}. Initially, both approaches obtain a reduction in $\Delta F$ with log-log slope of $-1/2$ [i.e., $\Delta F \propto (\textrm{time})^{-1/2}$], until converging to final values. 
Linear least-squares fits to the first five and seven points of the slice and pCN curves, respectively, give a $\sim$3.5$\times$ temporal speedup for pCN over slice at the same convergence level. Such an improvement---even for this comparatively small system of two qubits-- is significant for practical QST, where computational time represents an precious commodity. 

In the second test, we shift focus away from the sampling procedure and concentrate on the likelihood, comparing the full [Eq.~(\ref{e9})] and pseudo [Eq.~(\ref{e7})] versions directly. Since the experiment in question measured in combinations of the Pauli-$X$ and $Z$ bases for the two qubits, but not Pauli-$Y$, the experimental LS estimate consists of only eight of the fifteen total Pauli basis components, thus requiring the projector formalism in Fig.~\ref{fig1}(b). From a computational perspective, this projection can be efficiently implemented as a linear transformation,  by writing the density matrix elements as a length-$D^2$ column vector $\boldsymbol{\rho}_\mathrm{vec}$ and finding the matrix $V$ such that $\left[P_\mathcal{M}(\rho)\right]_\mathrm{vec} = V\boldsymbol{\rho}_\mathrm{vec}$, which can be precomputed according to the relationship between Pauli and computational basis representations~\cite{Gamel2016}. Likewise, the probabilities appearing in the full likelihood [Eq.~(\ref{e9})] can be vectorized so that $\log L_\bbD = \mathbf{N}^T \log W \boldsymbol{\rho}_\mathrm{vec}$, where $\mathbf{N}$ denotes the vector of counts and $W$ is the linear transformation mapping matrix elements to probabilities. Matrix-vector multiplication reduces function evaluation time and is essential in providing a fair comparison between the likelihood approaches.

\begin{figure*}[tb!]
\centering\includegraphics[width=2\columnwidth]{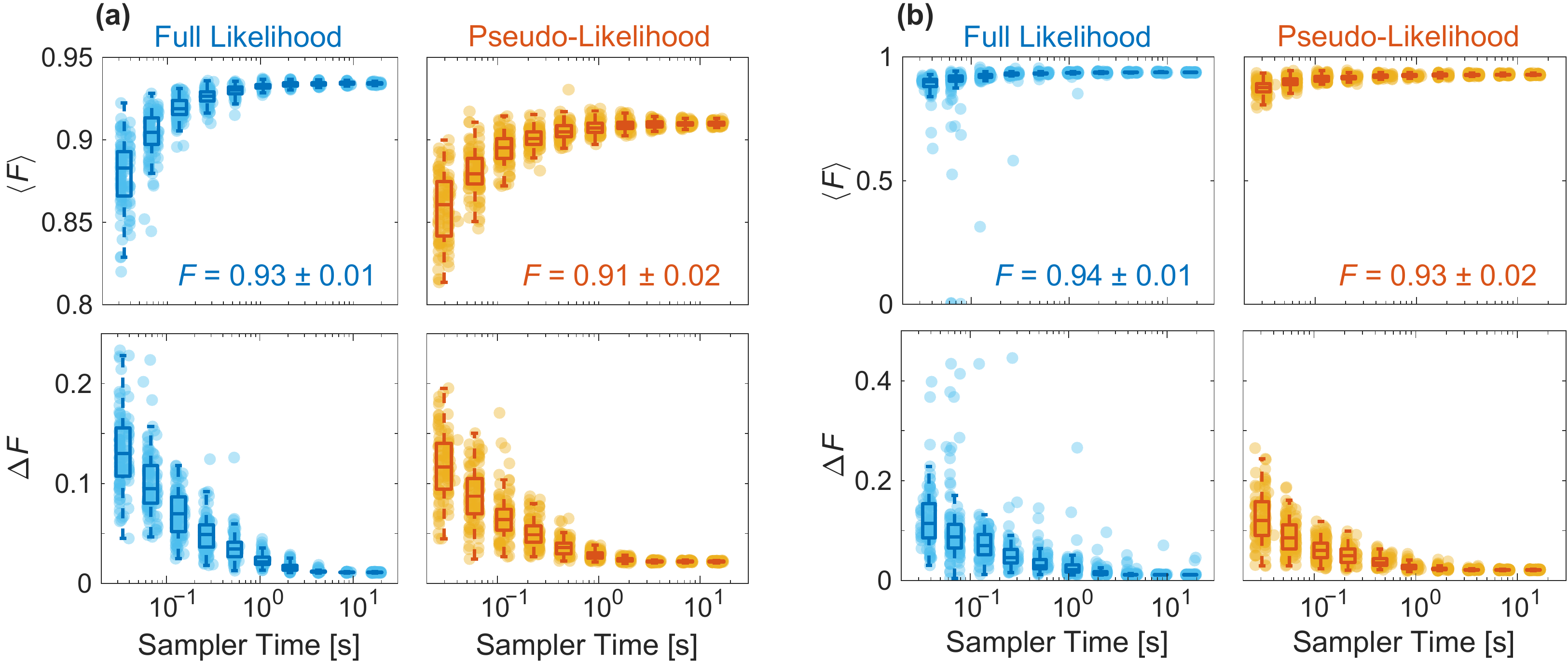}
\caption{Convergence plots for the priors (a) $\alpha=1$ and (b) $\alpha=1/4$, utilizing 100 independent pCN samplers for increasing Markov chain lengths.}
\label{fig4}
\end{figure*}

Figure~\ref{fig4}(a) plots $\braket{F}$ and $\Delta F$ for 100 samplers utilizing the pCN algorithm, as the number of points increases, for both the full (left) and psuedo (right) likelihood models; both consider $\alpha=1$ in the prior. [The full likelihood results are the same as Fig.~\ref{fig2}(b), reproduced here for comparison.] Both likelihoods converge to similar values, though the pseudo case returns slightly lower mean and higher uncertainty. The general congruity between the two cases offers evidence in favor of our choice for the variance ($\sigma^2 = 1/N$) and the projector-based approach to limited measurements. Even the slightly lower fidelity for the pseudo-likelihood is a positive feature, in that it does not overestimate the state's fidelity beyond that predicted by a complete model.

One of the advantages of the prior formulation is its ability to impart sparsity to the density matrix parameterization, through $\alpha$, and thus favor pure states. In Fig.~\ref{fig4}(b), we consider $\alpha=1/4$ and repeat the convergence tests with all other settings the same. In both the full and pseudo cases, the fidelity is slightly higher compared to $\alpha=1$, which makes sense in light of the extra weight given to pure states. Interestingly, the full-likelihood case shows additional outliers at this $\alpha$ value (note the much wider $y$-axis scale). Evidently, the sparser prior increases the tendency for trapping of the Markov chain around local maxima. By contrast, the pseudo-likelihood results remain much more consistent throughout the convergence plot. While it would be unwise to infer too much from acknowledged outliers, the pseudo-likelihood approach nonetheless appears slightly more robust to fluctuations in the sampling algorithm, a valuable feature for Bayesian QST.

Yet the pseudo-likelihood does not lead to any observable speedup in sampler time. In this particular example ($D=4$ and $16$ total measurement outcomes), such a situation obtains, first, because computational cost is dominated not by calculating $L_\bbD(\bx)$ but rather constructing $\rho(\bx)$ [Eq.~(\ref{e5})] and, second, because of the additional cost of computing the projection $P_\mathcal{M}(\rho)$ in this limited measurement case. As dimension $D$ increases, though, the pseudo-likelihood's improved efficiency should ultimately surface, a question we address with simulated data in the next section.

\section{Example with Simulated Data}
\label{sec:sim}
In order to explore dimensionalities beyond that of the experimental data available to us, we next generate simulated tomographic data for entangled two-qudit states with the ground-truth density matrix
\begin{equation}
\label{e10}
\rho = \lambda\ket{\Psi}\bra{\Psi} + \frac{1-\lambda}{D} I_D,
\end{equation}
where $\ket{\Psi}=\sum_{k=1}^{d}\ket{k}_A \ket{k}_B$ is a  high-dimensional Bell state, $D=d^2$, and $I_D$ is the the $D\times D$ identity matrix; $\lambda\in(0,1)$ controls the fidelity with respect to the ideal $\ket{\Psi}$. Count data is obtained by cycling through all pairwise combinations of $(d+1)$ MUBs~\cite{Wootters1989}, computing the $d^2$ outcome probabilities associated with the state in Eq.~(\ref{e10}), and drawing from a multinomial to emulate an experimental coincidence distribution. This procedure amounts to $Q=(d+1)^2 = (D+2\sqrt{D}+1)$ total measurement settings, each with $S_q = D$ outcomes, which are then either used to compute $\rho_{LS}$ and perform pseudo-likelihood--based QST or inserted directly as exponents in the full likelihood.

Explicitly, we consider qudit dimensions $d\in \{2,3,5,7 \}$ (Hilbert space dimensions $D\in \{4,9,25,49 \}$). Prime $d$ are chosen for convenience, for a complete $(d+1)$ set of MUBs can be generated easily in these cases utilizing Weyl operators~\cite{Sheridan2010}. We then set $\lambda=0.95$ for all tests and acquire $100D$ coincidences for each pair of bases in the simulated experiments. Running the pCN algorithm on these observations and recording the time per sample for $2^{14}$ points (following a burn-in period of $2^{10}$ points), we find the trends in Fig.~\ref{fig5}. While comparable at low $D$, the evaluation times for the full and pseudo-likelihood approaches become increasingly disparate as $D$ grows, reaching $\sim$10$\times$ for $D=49$. Due to limits on the size of the datasets we could generate, we were not able to reach the $\mathcal{O}(D^2)$ asymptotic scaling improvement of Sec.~\ref{sec:method}. Nonetheless, these time tests confirm that the pseudo-likelihood offers computational speedups under appropriate conditions.

\section{Conclusion}
Continued research should enable even further improvements to the sampling algorithm. While here we have applied the pCN approach to parameters with Gaussian prior distributions, pCN can be extended to non-Gaussian priors as well~\cite{vollmer}, provided one can select a proposal distribution which preserves reversibility with respect to the prior. It is worth noting also that there exist additional enhancements, for example utilizing derivative information, which are out of scope of the present work~\cite{law1, law1a, beskos1}. The method can also be embedded within SMC samplers~\cite{beskos2, law2}, and as $D\rightarrow \infty$ one can leverage finite approximations to further improve complexity~\cite{law2}. These directions are under investigation and will be reported in future work.

\begin{figure}[tb!]
\centering\includegraphics[width=\columnwidth]{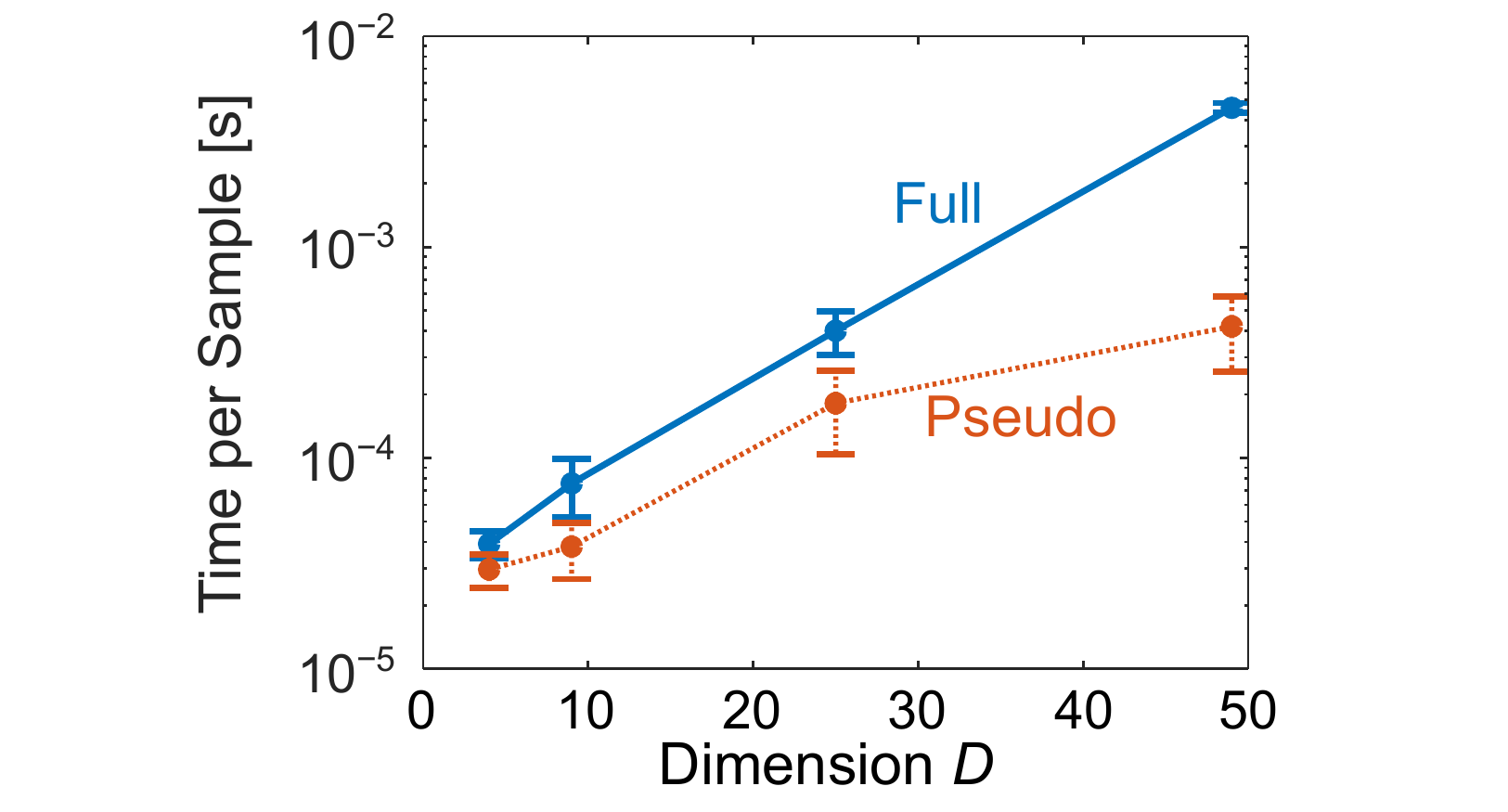}
\caption{Computational cost of our algorithm for simulated data with increasing dimension. Circles indicate the mean, and error bars the standard deviation, for $2^{14}$ points.}
\label{fig5}
\end{figure}

As it stands, the key features of the present method should find application in a myriad of quantum inference problems. With only minor modifications, models previously tackled by slice sampling---such as photon loss~\cite{Williams2017} or linear-optic transformations with dark counts~\cite{Lu2019a}---can be transformed into a pCN formalism for algorithmic speedup. On the other hand, the impact of the pseudo-likelihood in improving practical QST is less clear in our view. The pseudo-likelihood certainly appears more robust to initial conditions of the sampler [Fig.~\ref{fig4}(b)], with computational improvements for sufficiently large Hilbert spaces and complete measurements [Fig.~\ref{fig5}]. Yet at the dimensionalities where the pseudo-likelihood provides order-of-magnitude speedups (e.g., $D\sim 50$ in the example of Fig.~\ref{fig5}), it is possible that the \emph{experimental} challenge of acquiring the needed QST data will so outweigh the \emph{computational} challenge of evaluating the full likelihood as to render the pseudo-likelihood superfluous. That being said, the pseudo-likelihood's general, model-independent form could provide advantages which may not be evident in the specific QST problem of interest here, so that the full potential of the pseudo-likelihood remains an unanswered question.

In conclusion, we have introduced a Bayesian inference method for efficient quantum state tomography. Compatible with any number of observations, our approach enjoys all the standard advantages of Bayesian QST but with significantly improved computational efficiency, through a combination of well-chosen parameterization, likelihood, and MCMC sampling algorithm. Our numerical investigations on both real and simulated data confirm the promise of our approach, particularly the power of advanced statistical techniques such as pCN in practical quantum tomography.

\section*{Acknowledgments}
We thank R.~S. Bennink and B.~P. Williams for discussions. This work was funded by the U.S. Department of Energy, Office of Advanced Scientific Computing Research, through the Quantum Algorithm Teams and Early Career Research Programs. This work was performed in part at Oak Ridge National Laboratory, operated by UT-Battelle for the U.S. Department
of Energy under contract no. DE-AC05-00OR22725.


%

\end{document}